\documentclass[9pt,twocolumn,twoside]{opticajnl}
\journal{opticajournal} % use for journal or Optica Open submissions

\setboolean{shortarticle}{true}
\usepackage{mathrsfs}

\usepackage{lineno}
\title{Low noise resonant amplification by optical injection-locking and residual phase noise cancellation}

\author[1,*]{Y. Lange Simmons}
\author[1]{James Greenberg}
\author[1]{Brendan M. Heffernan}
\author[1]{Antoine Rolland}

\affil[1]{Boulder Research Labs, IMRA America, Inc., 1551 S Sunset St. Suite C, Longmont, CO, USA}

\affil[*]{lsimmons@imra.com}

\begin{abstract}
We demonstrate a low noise, high-gain, resonant optical amplifier that combines injection locking with feed-forward cancellation of residual phase noise. The wavelength-agnostic architecture uses a commercial semiconductor diode laser as a power amplifier while preserving the spectral purity of a weak reference. Although injection locking enforces phase coherence, finite residual phase noise within the locking regime limits high-fidelity transfer of low phase noise from the reference laser to the injection-locked laser, particularly at large gain. Here, the residual phase error is measured via optical heterodyne detection and canceled using feed-forward phase correction. Compared to injection locking alone, the amplifier achieves up to 38~dB phase-noise reduction at Fourier frequencies above 1~kHz for injection ratios down to -57~dB. This approach enables ASE-free amplification of low-power, low-noise optical references, including individual lines from optical frequency combs.
\end{abstract}

\setboolean{displaycopyright}{false} % Do not include copyright or licensing information in submission.

\begin{document}

\maketitle

%Alternative intro
Narrow-linewidth, low phase-noise lasers are essential for photonic microwave generation, where optical coherence is directly mapped to low-noise electrical signals~\cite{xie_photonic_2017, liu_photonic_2020}, for optical frequency transfer, where phase noise limits long-distance stability~\cite{hoghooghi_ultrastable_2025}, for coherent atomic control, where laser noise directly degrades transition fidelity~\cite{day_limits_2022,krinner_low_2024}, and for advanced optical communication systems, where phase noise constrains coherent detection and spectral efficiency~\cite{fang_overcoming_2024, chang_modeling_2025}. In many of these applications, the available optical reference is intrinsically weak, such as an individual line from an optical frequency comb or a heavily attenuated signal at a fiber repeater station, while milliwatt-level optical power is required downstream. Achieving large optical gain while preserving spectral purity in this small-signal regime remains a central challenge.

Conventional optical amplifiers are poorly suited to this task. Erbium-doped fiber amplifiers (EDFA's) and semiconductor optical amplifiers (SOA's) provide high gain but exhibit poor noise performance for weak input signals, where amplified spontaneous emission (ASE) raises the noise floor and produces a broadband background that often necessitates aggressive spectral filtering~\cite{mikitchuk_noise_2017,desurvire_erbium-doped_2002, connelly_semiconductor_2017, pedersen_small-signal_1994}. In practice, ultra-narrow optical filters increase system complexity and loss while failing to fully recover the spectral purity of the input signal. As a result, EDFAs and SOAs fundamentally struggle to amplify weak, single-frequency optical fields without degrading coherence.

Injection locking offers an attractive alternative by simultaneously providing amplification and coherent spectral transfer without ASE. However, achieving a sufficiently wide injection bandwidth requires the use of low-\(Q\), broad-linewidth diode lasers. Because injection locking behaves as a first-order (type-I) phase transfer process, suppression of the diode phase noise is limited to a single-pole response, leading to only partial rejection of phase fluctuations within the locking bandwidth.~\cite{gardner_phaselock_2005,chow_phase_1982, heng-chia_chang_phase_1997, razavi_study_2004}. Consequently, even with broadband injection, residual phase noise from the injection-locked laser can dominate at high Fourier frequencies and limit the fidelity of spectral purity transfer from an ultra-low-noise reference.

Here we present a low-noise optical amplifier that combines the complementary strengths of injection locking and feed-forward phase noise cancellation. Injection locking provides ASE-free amplification and coarse spectral transfer, while a feed-forward correction explicitly measures and cancels the residual phase noise introduced by the injection-locked oscillator~\cite{chao_pounddreverhall_2024, li_active_2022}. This architecture enables high optical gain from extremely weak inputs without requiring narrow spectral filtering. We demonstrate operation with input powers as low as 100~nW (injection ratios less than -55~dB), and phase-noise reductions greater than 30~dB from 1~kHz to 100~kHz Fourier frequencies. The amplifier is well suited for applications requiring coherent regeneration of weak optical signals, including optical frequency transfer, clock distribution, and photonic microwave and terahertz generation.

\begin{figure*}[tb]
\centering
{\includegraphics[width=0.8\linewidth]{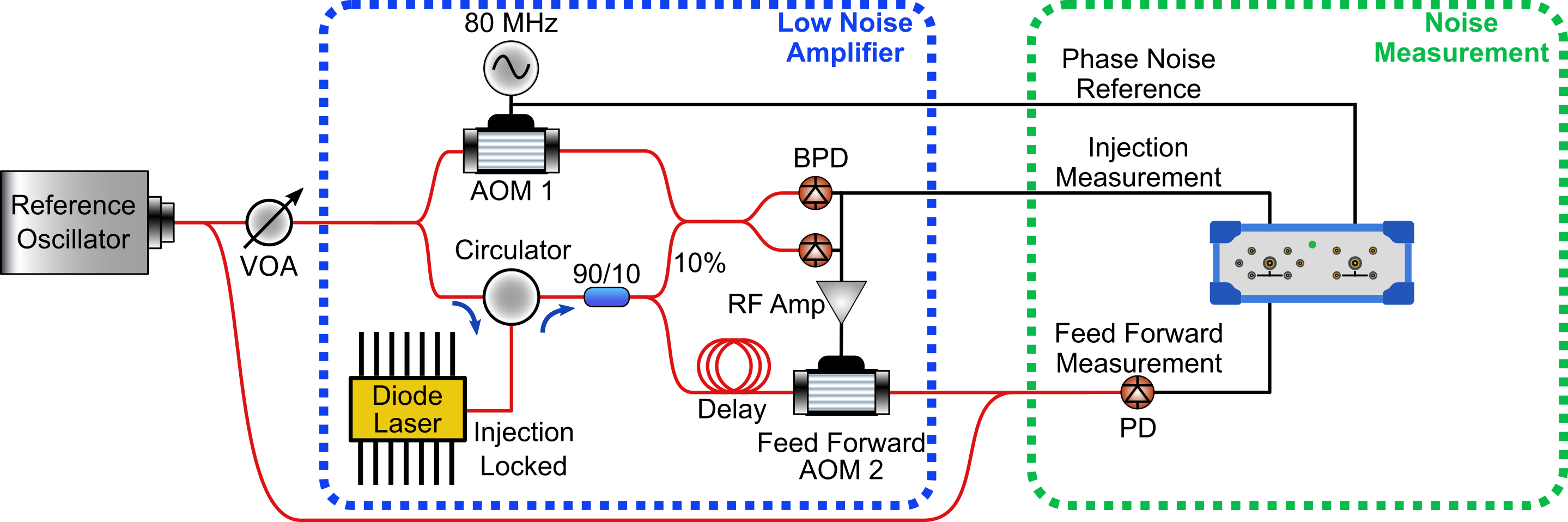}}
\caption{Schematic of the low-noise optical amplifier and phase-noise measurement setup. Light from a low-noise reference laser is attenuated and split, with one arm injected into a high-power diode laser through an optical circulator to achieve injection locking. The remaining reference light is frequency shifted by AOM 1 and heterodyned with a fraction of the injection-locked output on a balanced photodiode (BPD) to generate a feed-forward error signal carrying the residual phase noise of injection. This signal drives a second AOM to apply feed-forward phase correction to the amplified output. An optional fiber delay compensates the latency between heterodyne detection and correction. The corrected output is heterodyned with the reference laser for residual phase-noise measurement.}
\label{fig:system-diagram}
\end{figure*}

\paragraph{Theoretical Overview.}
The first stage of the amplifier employs optical injection locking, in which light from a low-noise reference laser is injected into a higher-power diode laser. When the reference frequency lies within the locking bandwidth, the diode laser follows the reference in both frequency and phase. The locking bandwidth ($\Delta f_{\mathrm{lock}}$) is given by
\begin{equation}
\Delta f_{\mathrm{lock}} = \rho\,f_{3\mathrm{dB}},\label{eq:lockrange}
\end{equation}
where $f_{3\mathrm{dB}}$ is the intrinsic 3-dB bandwidth of the diode laser and $\rho = \sqrt{P_{\mathrm{inj}}/P_{\mathrm{out}}}$ is the injection ratio~\cite{liu_optical_2020,heng-chia_chang_phase_1997}.

We denote by $\varphi(t)$ the phase fluctuations of an optical field and by $S_{\varphi}(f)$ the corresponding phase-noise power spectral density.

Injection locking can be interpreted as a first-order (type-I) phase-locked loop, in which the injected field provides a proportional phase error signal~\cite{buczek_laser_1973,gardner_phaselock_2005,razavi_study_2004}. The phase transfer therefore exhibits finite loop gain, and the injection-locked phase can be written as
\begin{equation}
\varphi_{\mathrm{inj}}(t) = \varphi_{\mathrm{reference}}(t) + \varphi_{\mathrm{residual}}(t),
\end{equation}
where $\varphi_{\mathrm{residual}}(t)$ represents the intrinsic locked-state phase error.

From a control perspective, the reference phase is transferred through a low-pass response, while the residual phase follows the complementary high-pass response. The corresponding transfer function is
\begin{equation}
|H_{\mathrm{res}}(f)|^2 \approx \frac{f^2}{f^2 + \Delta f_{\mathrm{lock}}^2}.
\end{equation}
This leads to a residual phase-noise PSD
\begin{equation}
S_{\varphi,\mathrm{residual}}(f) = |H_{\mathrm{res}}(f)|^2\,S_{\varphi,\mathrm{free}}(f),
\end{equation}
where $S_{\varphi,\mathrm{free}}(f)$ is the phase-noise PSD of the free-running diode laser. As $\Delta f_{\mathrm{lock}}$ decreases, the residual phase noise increases, fundamentally limiting the transfer of spectral purity.

The residual phase error can be measured via optical heterodyne detection. The beat-note phase fluctuations are
\begin{equation}
\varphi_{\mathrm{beat}}(t) = \varphi_{\mathrm{RF}}(t) + \varphi_{\mathrm{residual}}(t),
\label{eq:phi_beat}
\end{equation}
where $\varphi_{\mathrm{RF}}(t)$ is the phase fluctuation introduced by the RF frequency shifter.

Feed-forward cancellation is implemented using an acousto-optic modulator (AOM) driven by the heterodyne signal. The output phase is
\begin{equation}
\varphi_{\mathrm{out}}(t) = \varphi_{\mathrm{reference}}(t) + \varphi_{\mathrm{residual}}(t)
\pm \left[\varphi_{\mathrm{RF}}(t) + \varphi_{\mathrm{residual}}(t)\right].
\end{equation}
Selecting the $-1$\textsuperscript{st} order cancels the residual phase terms, yielding
\begin{equation}
\varphi_{-1,\mathrm{out}}(t) = \varphi_{\mathrm{reference}}(t) - \varphi_{\mathrm{RF}}(t),
\end{equation}
so that the residual phase error of injection locking is explicitly removed. The corresponding phase-noise PSD is then determined by the reference and the RF source.

A practical limitation arises from the finite delay $\tau$ between detection and correction. This introduces a residual phase-noise contribution
\begin{equation}
S_{\varphi,\mathrm{delay}}(f) \approx S_{\varphi,\mathrm{residual}}(f)\,(2\pi f\tau)^2,
\label{eq:delay-impact}
\end{equation}
which increases with Fourier frequency and ultimately limits high-frequency noise suppression.

\paragraph{Experimental Implementation.} A schematic overview of the low-noise amplifier and its characterization is shown in Fig.~\ref{fig:system-diagram}. A low-noise stimulated Brillouin scattering (SBS) laser was used as the optical reference source~\cite{heffernan_brillouin_2024}. The optical power delivered to the amplifier was controlled using a variable optical attenuator (VOA). The reference light was split by a fiber coupler, with one path used for heterodyne feed-forward and the other injected into the diode laser via an optical circulator. The diode laser was a high-power, fiber-coupled, single-mode device (EP1550-ADF, Eblana) with a side-mode suppression ratio of approximately $-45~\mathrm{dB}$ and a nominal output power of $40~\mathrm{mW}$. Its drive current and temperature were adjusted to tune the free-running frequency, ensuring operation within the injection-locking bandwidth. %The experimental realization of the amplifier used a low-noise stimulated Brillouin scattering (SBS) laser as the optical reference source, as described in~\cite{heffernan_brillouin_2024}. The SBS laser also served as the reference for measuring the residual phase noise of the amplifier. As shown in Fig.~\ref{fig:system-diagram}, the low-noise amplifier consists of an injection-locked diode laser, an optical heterodyne measurement of the residual phase error, and a feed-forward phase noise correction stage.

\begin{figure*}[tb]
% \fbox{\includegraphics[width=\linewidth]{PN_compare_test_text.pdf}}
{\includegraphics[width=0.95\linewidth]{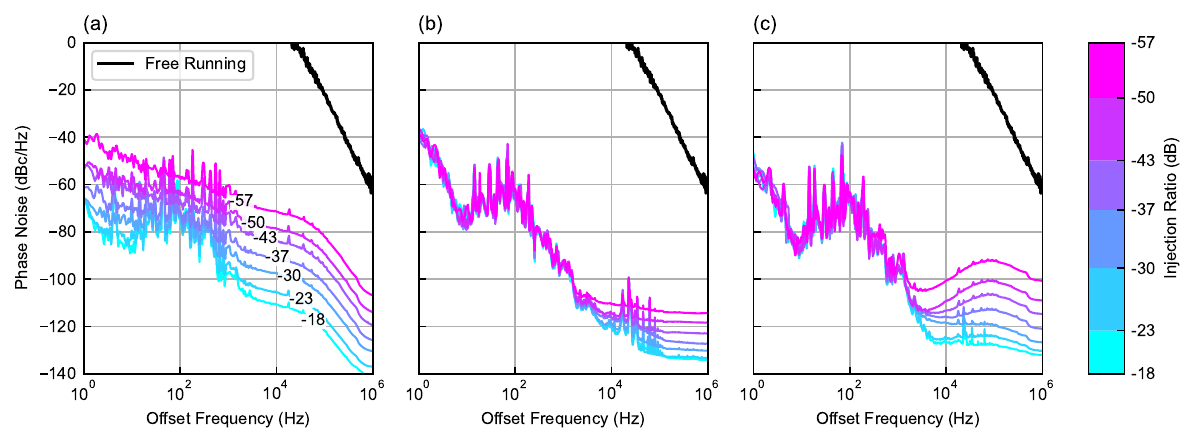}}
\caption{Residual phase noise comparison of (a) the injection-locked oscillator alone, (b) including feed-forward correction and delay line compensation, (c) and feed-forward without delay line compensation. Each trace is labeled in (a) with the injection ratio in dB units. Free running noise of the laser diode is included in each plot.}
\label{fig:PN-compare}
\end{figure*}

%The optical power delivered to the amplifier was controlled using a variable optical attenuator (VOA). The reference light was split by a fiber coupler, with one path injected into the injection-locked laser and the other used to generate the heterodyne signal for feed-forward correction. Injection was implemented using an optical circulator and a high-power, fiber-coupled, single-mode diode laser (EP1550-ADF, Eblana) with a side-mode suppression ratio of approximately -45~dB. The drive current and temperature of the injection-locked laser were adjusted to tune its free-running frequency, ensuring operation within the injection-locking bandwidth. The nominal output power of the injection-locked laser was 40~mW.

The feed-forward error signal was generated using an optical heterodyne measurement. A portion of the reference light not used for injection was frequency shifted by $80~\mathrm{MHz}$ using an acousto-optic modulator (AOM, T-M080, Gooch \& Housego) driven by a low-noise RF oscillator. The shifted light was combined with 10\% of the diode laser output and detected using a balanced photodiode (BPD). The resulting beat note carries the residual phase fluctuations of the injection-locked laser, as described by~\eqref{eq:phi_beat}.

Feed-forward correction was implemented using a second AOM placed in the diode laser output path. To compensate for the delay between heterodyne detection and phase modulation, an optional fiber delay line was inserted before the correction AOM. A length of $100~\mathrm{m}$ of single-mode fiber was used to delay the optical signal, partially compensating the feed-forward latency, which is dominated by the AOM response time~\cite{zhang_linewidth_2016}.

\begin{figure}[tb]
\centering
{\includegraphics[width=0.95\linewidth]{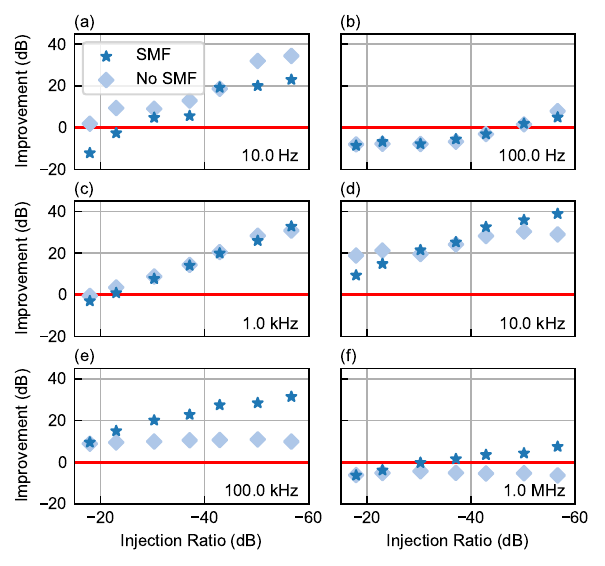}}
\caption{The change of residual phase noise versus injection ratio (horizontal axis inverted) of the feed-forward amplifier compared to injection-locking alone. Values are displayed for amplification with and without single mode fiber (SMF) delay compensation. Comparisons are performed at (a) 10~Hz, (b) 100~Hz, (c) 1~kHz, (d) 10~kHz, (e) 100~kHz, and (f) 1~MHz Fourier frequency.}
\label{fig:improvements}
\end{figure}

Residual phase-noise measurements were performed to quantify the amplifier performance and the improvement provided by feed-forward correction. Figure~\ref{fig:PN-compare}a shows the residual phase noise of the injection-locked diode as a function of injection ratio, spanning values from $-57$ to $-18~\mathrm{dB}$. The data were obtained directly from the feed-forward error signal, analyzed using a phase-noise analyzer (53100A, Microchip) referenced to the RF drive used for the optical heterodyne measurement~\cite{teyssieux_phase_2026}. Increasing injection power reduces phase noise at low Fourier frequencies, consistent with an increase in the effective locking bandwidth. The intrinsic $f_{3\mathrm{dB}}$ bandwidth of the diode was determined to be $77~\mathrm{GHz}$ by measuring the injection-locking bandwidth at multiple injection powers and fitting the dependence using~\eqref{eq:lockrange}.

The free-running phase noise of the diode laser is also shown for comparison. Owing to large frequency drifts in free-running operation, conventional phase-noise analysis was not feasible. Instead, the noise was characterized from the quadrature components of the optical heterodyne beat note with the SBS reference laser using an RF signal analyzer, following~\cite{schiemangk_accurate_2014}.

Figure~\ref{fig:PN-compare}b shows the phase noise of the full amplifier after feed-forward correction with delay compensation. The measurement was performed by heterodyning the amplifier output with the SBS reference, taken before the VOA in Fig.~\ref{fig:system-diagram}. A clear reduction in phase noise is observed above $10~\mathrm{kHz}$, where residual injection-locking noise is otherwise dominant.

At Fourier frequencies between $10~\mathrm{Hz}$ and $1~\mathrm{kHz}$, the phase noise is limited by technical contributions, including environmental perturbations and added fiber noise. Figure~\ref{fig:PN-compare}c shows the feed-forward–corrected phase noise measured without the $100~\mathrm{m}$ delay fiber. In this case, high-frequency suppression is reduced due to imperfect delay matching, while performance at $1$--$10~\mathrm{Hz}$ is improved. This behavior reflects the reduced sensitivity of feed-forward correction to delay mismatch at low Fourier frequencies, as well as the removal of thermal and acoustic noise introduced by the additional fiber length.

\begin{figure}[!h]
\centering
{\includegraphics[width=0.95\linewidth]{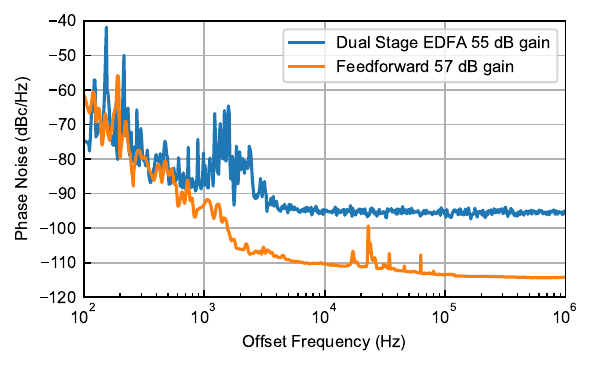}}
\caption{Residual phase noise comparison of the low noise amplifier and a dual stage Erbium-dopde fiber amplifier (EDFA), which included spectral filtering to reduce ASE. The total EDFA gain was 55~dB and the low noise amplifier injection ratio shown is -57~dB (corresponding to a gain of 57 dB).}
\label{fig:EDFA-compare}
\end{figure}

% Residual phase noise of the injection-locked diode is measured before the feed-forward process, using the feed-forward signal. An identical phase noise measurement was carried out for the amplified, feed-forward corrected signal. The amplified light was optically heterodyned with light from the SBS laser, from a tap before the VOA and detected with a balanced photodiode (PDM480C-AC, Thorlabs). Phase noise measurements were acquired for both signals with a phase noise analyzer , using the 80 MHz RF driving the optical heterodyne AOM as a reference. Figure \ref{fig:PN-compare} shows the phase noise measurements, representing the residual, or added, phase noise onto the reference source. % An intensity noise analyzer (PNA1, Thorlabs) measured the relative intensity noise (RIN). 

% Phase noise measurements were acquired for a range of input powers, defined as the power sent to injected-laser after attenuation and losses, with equal power sent through the optical heterodyne path (Fig. \ref{fig:system-diagram}).  The residual phase noise, $\mathscr{L}(f)$ in units of $\mathrm{dBc/Hz}$, is measured using the RF driving frequency as the reference.~\cite{teyssieux_phase_2025}

Comparison of the phase-noise data in Fig.~\ref{fig:improvements} highlights the dependence of noise suppression on both injection ratio and Fourier frequency. The largest improvements occur at low injection ratios, where the effective locking bandwidth is reduced and the residual phase noise is correspondingly increased. In this regime, feed-forward correction provides substantial suppression of the residual phase error that cannot be mitigated by injection locking alone.

% Phase noise improvement is observed at 100~Hz and at offset frequencies above approximately 10~kHz. At low offset frequencies below 1~kHz, the measured performance is limited by excess noise introduced by the additional fiber delay, including thermal and acoustic perturbations. At higher offset frequencies, the benefit of feed-forward correction increases as residual phase noise becomes the dominant limitation.

At an injection ratio of $-57~\mathrm{dB}$, the phase noise at a $1~\mathrm{MHz}$ offset is improved by $7.4~\mathrm{dB}$ with delay compensation, but degraded by $6.4~\mathrm{dB}$ without it. This highlights the critical role of delay matching for high-frequency noise suppression, consistent with the scaling predicted by~\eqref{eq:delay-impact}.

In addition to delay-related limitations, the feed-forward architecture introduces an additive noise floor at high Fourier frequencies, resulting in an increase of the white phase-noise level when the correction path is active. This behavior arises from noise contributions in the correction chain, including the finite signal-to-noise ratio of the optical heterodyne measurement, noise added by the RF drive electronics, and reduced optical power in the corrected first diffracted order. These contributions are not correlated with the residual phase fluctuations of the injection-locked laser and therefore cannot be suppressed by feed-forward correction.

This behavior is analogous to a noise figure for phase noise: while the feed-forward process suppresses correlated phase fluctuations, it introduces uncorrelated noise that sets a floor for the achievable performance. Consistent with this interpretation, the high-frequency noise floor depends on the optical power of the reference signal. Lower reference power reduces the heterodyne signal-to-noise ratio, increasing the phase noise of the error signal and leading to a corresponding degradation of the corrected output.

For comparison, the amplifier was benchmarked against an erbium-doped fiber amplifier (EDFA). Two EDFA stages were used to achieve a total gain comparable to that of the low-noise amplifier. The attenuated SBS source at $1549.6~\mathrm{nm}$ was first amplified and then spectrally filtered using a tunable optical filter to suppress amplified spontaneous emission (ASE). A second EDFA stage increased the optical power to $40~\mathrm{mW}$, matching the output power of the diode laser prior to feed-forward correction.

Figure~\ref{fig:EDFA-compare} compares the residual phase noise of the two-stage EDFA, providing $55~\mathrm{dB}$ of gain, with that of the low-noise amplifier operating at an injection ratio of $-57~\mathrm{dB}$ (corresponding to a gain of $57~\mathrm{dB}$). Despite spectral filtering, the EDFA exhibits an elevated white phase-noise floor near $-95~\mathrm{dBc/Hz}$. This degradation arises from amplified spontaneous emission (ASE) associated with high gain applied to a low-power ($\sim\!100~\mathrm{nW}$) input signal. In contrast, the injection-locked, feed-forward–corrected amplifier preserves low phase noise without introducing ASE, demonstrating a clear advantage for amplifying weak, spectrally pure optical references~\cite{mikitchuk_noise_2017}.

In conclusion, we have demonstrated a high-gain, low-noise optical amplifier combining injection locking with feed-forward cancellation of residual phase noise. By explicitly correcting the finite locked-state phase error inherent to injection locking, this approach enables ASE-free amplification of nanowatt optical references with gains exceeding $50~\mathrm{dB}$. Compared to injection locking alone, phase noise is reduced by up to $38~\mathrm{dB}$ at Fourier frequencies above $10~\mathrm{kHz}$, with performance ultimately limited by delay matching in the feed-forward path.

More broadly, this work highlights a fundamental limitation of injection locking when viewed as a type-I phase-locked loop, namely the incomplete transfer of phase stability at high gain. The feed-forward architecture provides a general and modular route to overcoming this limitation without requiring high-bandwidth feedback or optical cavities.

The demonstrated amplifier is directly applicable to the amplification of individual lines from optical frequency combs, low-noise laser dissemination, and photonic microwave and terahertz generation, where preservation of spectral purity at low optical power is critical. Future implementations may benefit from integrated modulators and reduced latency to extend noise suppression to higher Fourier frequencies. More broadly, the combination of injection locking and feed-forward correction provides a scalable framework for precision optical amplification in regimes where conventional gain media introduce prohibitive noise.

%\author{Author One\authormark{1} and Author Two\authormark{2,*}}

%\address{\authormark{1}Peer Review, Publications Department,
%Optica Publishing Group, 2010 Massachusetts Avenue NW,
%Washington, DC 20036, USA\\
%\authormark{2}Publications Department, Optica Publishing Group,
%2010 Massachusetts Avenue NW, Washington, DC 20036, USA\\
%%\authormark{3}xyz@optica.org}

%\email{\authormark{*}xyz@optica.org}}

%Example with the corresponding author designated by an asterisk and a note indicating equal contributions by two authors.

%\author{Author One\authormark{1,3} and Author %Two\authormark{2,3,*}}

%\address{\authormark{1}Peer Review, Publications Department,
%Optica Publishing Group, 2010 Massachusetts Avenue NW, %Washington, DC 20036, USA\\
%\authormark{2}Publications Department, Optica Publishing Group, %2010 Massachusetts Avenue NW, Washington, DC 20036, USA\\
%\authormark{3}The authors contributed equally to this work.\\
%\authormark{*}xyz@optica.org}}

%\section{Examples of Article Components}
%\label{sec:examples}

\begin{backmatter}

% \bmsection{Acknowledgment} 

% \smallskip

% \noindent Here are examples of disclosures:

\bmsection{Disclosures} The authors declare no conflicts of interest.

\bmsection{Data availability} Data underlying the results presented in this paper are not publicly available at this time but may be obtained from the authors upon reasonable request.

\end{backmatter}

% Bibliography
\bibliography{library}

% Full bibliography added automatically for Optics Letters submissions; the following line will simply be ignored if submitting to other journals.
% Note that this extra page will not count against page length
\bibliographyfullrefs{library}

% Please include bios and photos of all authors for aop articles
\ifthenelse{\equal{\journalref}{aop}}{%
\section*{Author Biographies}
\begingroup
\setlength\intextsep{0pt}
\begin{minipage}[t][6.3cm][t]{1.0\textwidth} % Adjust height [6.3cm] as required for separation of bio photos.
  \begin{wrapfigure}{L}{0.25\textwidth}
    \includegraphics[width=0.25\textwidth]{john_smith.eps}
  \end{wrapfigure}
  \noindent
  {\bfseries John Smith} received his BSc (Mathematics) in 2000 from The University of Maryland. His research interests include lasers and optics.
\end{minipage}
\begin{minipage}{1.0\textwidth}
  \begin{wrapfigure}{L}{0.25\textwidth}
    \includegraphics[width=0.25\textwidth]{alice_smith.eps}
  \end{wrapfigure}
  \noindent
  {\bfseries Alice Smith} also received her BSc (Mathematics) in 2000 from The University of Maryland. Her research interests also include lasers and optics.
\end{minipage}
\endgroup
}{}

\end{document}